\documentclass{article}
\usepackage[utf8]{inputenc}
\usepackage{amsmath,bm,graphicx,url}
\usepackage{color, colortbl}
\definecolor{highlight}{rgb}{1,.88,.88}

\title{Modeling strict age-targeted mitigation strategies for COVID-19}
\author{Maria Chikina\thanks{Department of Computational and Systems Biology, University of Pittsburgh Medical School, 3501 Fifth Ave, Pittsburgh, PA 15213}~  and Wesley Pegden\thanks{Dept.~of Mathematical Sciences, Carnegie Mellon University, Pittsburgh PA 15213}}
\date{April 2020}

\usepackage{fullpage}

\usepackage{hyperref}
\hypersetup{colorlinks,linkcolor=black,filecolor=black,urlcolor=black,citecolor=black}

\newcommand{\ccc}{\mathcal{C}}
\newcommand{\mmm}{\mathcal{M}}
\newcommand{\iii}{\mathcal{I}}

\newcommand{\bdel}{{\bm{\delta}}}
\newcommand{\bgam}{{\bm{\gamma}}}
\newcommand{\brho}{{\bm{\rho}}}

\newcommand{\comment}[1]{}

\begin{document}

\maketitle

\begin{abstract}
We use a simple SIR-like epidemic model which integrates known age-contact patterns for the United States to model the effect of age-targeted mitigation strategies for a COVID-19-like epidemic.  We find that, among strategies which end with population immunity, strict age-targeted mitigation strategies have the potential to greatly reduce mortalities and ICU utilization for natural parameter choices.
\end{abstract}

\section{Introduction}
There is currently a shortage of publicly available COVID-19 modeling which considers strict but age-targeted mitigation strategies---in particular, strategies which may divide the labor force into low-risk and high-risk groups.

In this paper, we use a simple age-sensitive SIR (Susceptible, Infected, Removed) model integrating known age-interaction contact patterns to examine the potential effects of age-heterogeneous mitigations on an epidemic in a COVID-19-like parameter regime.  Our goal is to demonstrate the qualitative point that for an epidemic with COVID-19-like parameters, age-sensitive mitigations can result in considerably less mortality and ICU usage than homogeneous mitigations, among strategies which end with population immunity.

We model mitigations which result in a 70\% reduction in transmission rates among all of the population except for a relaxed group.  In our models, natural transmission rates correspond to an $R_0$ of 2.7.  (In Section \ref{s.sensitivity}, we consider other parameter choices.) We consider relaxation strategies based on age thresholds; for example, relaxing restrictions on all people under 50.  Because people of different ages often live in the same household, and because other risk factors might also be used to inform mitigation efforts, we consider strategies where only some fraction of people under the age threshold are subject to relaxed restrictions.

Related work has in the context of MERS can be found in \cite{thesis}, and for H1N1 swine flu in \cite{wildlife}.

\section{The model}
\label{themodel}
We use a simple SIR-like model acting on $n$ age groups.  Let $I(t),S(t),R(t),M(t)$ be vector-valued functions of time, each vector of length $n$, where the $j$th coordinate of each of $I(t),S(t),R(t),$ and $M(t)$, respectively, denote the the number of people in the $j$th age group who are infected, susceptible, recovered, or deceased at time $t$.

We let $\ccc$ be the age-group contact matrix.  This is a symmetric matrix describing the frequency of transmission events between age groups, relative to the products of the population sizes.  (In particular, even if the sizes of age groups vary, $\ccc$ would be a constant matrix if we assumed people were equally likely to have have transmission interactions regardless of age.)  The constant $\alpha$ is the total recovery/removal rate, and the vector $\bdel=(\delta_1,\dots,\delta_n)$ gives the death rate for each group separately.

Given a vector $v$, we let $\mmm_v$ denote the diagonal matrix whose $i$th diagonal entry is the $i$th coordinate of $v$.

%Finally, we let $\rho_1,\dots,\rho_n$ be the fractions of the total populations in age groups $1\dots n$ and let $[\rho_{i}]$ denote the diagonal matrix whose $i$th diagonal entry is $\rho_i$.

The dynamics of the model are captured by the following vector differential equations:
\begin{align}
\label{dIdt}\frac{dI(t)}{dt}&=I(t)\ccc \mmm_{S(t)}\cdot \frac{1}{m}-\alpha I(t)\\
\label{dSdt}\frac{dS(t)}{dt}&=-I(t)\ccc \mmm_{S(t)}\cdot \frac{1}{m} \\
\label{dRdt}\frac{dR(t)}{dt}&=(\alpha\mathbf{1}-\bdel)\odot I(t)\\
\label{dMdt}\frac{dM(t)}{dt}&=\bdel\odot I(t).
\end{align}
Here $\odot$ denotes coordinate-wise multiplication, and $m$ is the total population.

When modeling mitigation strategies, the contact matrix $\ccc$ will be modified.  In the absence of mitigations, we could simply set $\ccc$ to be a scalar multiple of the all-1's matrix $J$ (with the scalar set to reproduce the desired $R_0$ value).  More realistically, we can incorporate known patterns of inter-age-group interactions.  To do this, we use contact matrices generated by \cite{Contact}, see Section \ref{s.Contact}; these are are extrapolations based on contact patterns originally measured by recording contact patterns for a large sample of people in European countries.  In this manuscript, we use the contact matrices from \cite{Contact} generated for the United States.

One feature of using a non-constant contact matrix $\ccc$ is that the steady state proportions of a growing epidemic are no longer uniform among the age-groups in the population.  For example, if we model a situation where we begin with $10^5$ infections in a much larger population, we do not expect those to be uniformly distributed by age, but instead distributed in a way which depends on $\ccc$.

To understand the dependence on $\ccc$, let $\bgam$ denote a vector of length $n$ whose proportions represent infected fractions of each age group in the early growth of an epidemic (i.e., in the period where $R_0$ is often measured in the field).  We call this the \emph{early-stable proportion} vector.  We let $\brho=(\rho_1,\dots,\rho_n)$ be the vector giving the fraction of the total population within each age group.  If the \emph{relative} magnitudes of the coordinates of $\bgam$ are in steady-state (that is, if $\bgam$ will grow in all coordinates uniformly), this means from \eqref{dIdt}, that for some constant $K$ we have that
\[
\bgam\ccc\mmm_{\brho}-\alpha\bgam=K\bgam
\]
In particular, $\bgam$ is a positive eigenvector of the positive matrix $\ccc\mmm_{\brho}$.  Such an eigenvector exists and is unique, by the Perron–Frobenius theorem.  We let $\lambda_1$ denote its eigenvalue, which is the (unique) largest eigenvalue of $\ccc\mmm_\brho$.

To tie our model to empirical estimates of $R_0$ for COVID-19, we wish to compute a scaling factor $\beta$ for the matrix $\ccc$ which gives rise to an exponential growth in $\sum_{j=1}^n I_j(t)$ which corresponds to the empirical $R_0$ value computed early in an epidemic (before a substantial fraction of the population is infected), given an infection distributed according to the early-stable proportion vector $\bgam$.  For the vector $\bgam$, multiplication by $\ccc\mmm_\rho$ is equivalent to multiplication by the eigenvalue $\lambda_1$, thus for an infected population with age distribution proportional to $\bgam$, and writing $\iii$ for $\sum_{j=1}^n I_j$, the dynamics near $t=0$ reduce \eqref{dIdt} to a standard single population SIR model with the infected population governed by the equation
\[
\frac{d\iii(t)}{dt}=\frac{\lambda_1}{m} \iii(t)-\alpha \iii(t).
\]
In particular, we see that $R_0=\frac{\lambda_1}{\alpha}$, and thus the correct scaling for the contact matrix $\ccc$ to replicate the known $R_0$ value is that which ensures that \[
\lambda_1=R_0\cdot \alpha.
\]

We use the recovery rate $\alpha=1/14$, corresponding to a 2-week recovery period.  The mortality rate and rate of ICU admissions per infection are taken from Report 9 of the team at Imperial College London \cite{Imp9}; we use the data from their Table 3, which we reproduce here:

\begin{center}
\small
\begin{tabular}{c||c|c|c}
Age-group &\begin{tabular}{c}\% symptomatic cases \\ req.~hospitalisation\end{tabular}&\begin{tabular}{c}\% hospitalised cases\\ req.~critical care \end{tabular}& IFR\\
\hline
\hline
0 to 9  & 0.1\%& 5.0\%& 0.002\%\\
\hline
10 to 19& 0.3\%&5.0\%&0.006\% \\
\hline
20 to 29&1.2\%&5.0\%&0.03\% \\
\hline
30 to 39&3.2\%&5.0\%&0.08\%\\
\hline
40 to 49&4.9\%&6.3\%&0.15\%\\
\hline
50 to 59&10.2\%&12.2\%&0.60\%\\
\hline
60 to 69&16.6\%&27.4\%&2.2\%\\
\hline
70 to 79&24.3\%&43.2\%&5.1\%\\
\hline
80+&27.3\%&70.9\%&9.3\% 
\end{tabular}
\end{center}

Note that we do not apply a correction for asymptomatic cases, which may make our analysis conservative (pessimistic).  Comparable (but not identical) estimates may be found from other sources. For example, see \cite{LonTab}, Table S2.

\bigskip
\noindent Source code for our model can be found at the entry for this paper at \url{http://math.cmu.edu/~wes/pub.html}.

\section{The contact matrix}
\begin{figure}[h]
    \centering
    \includegraphics[width=.5\linewidth]{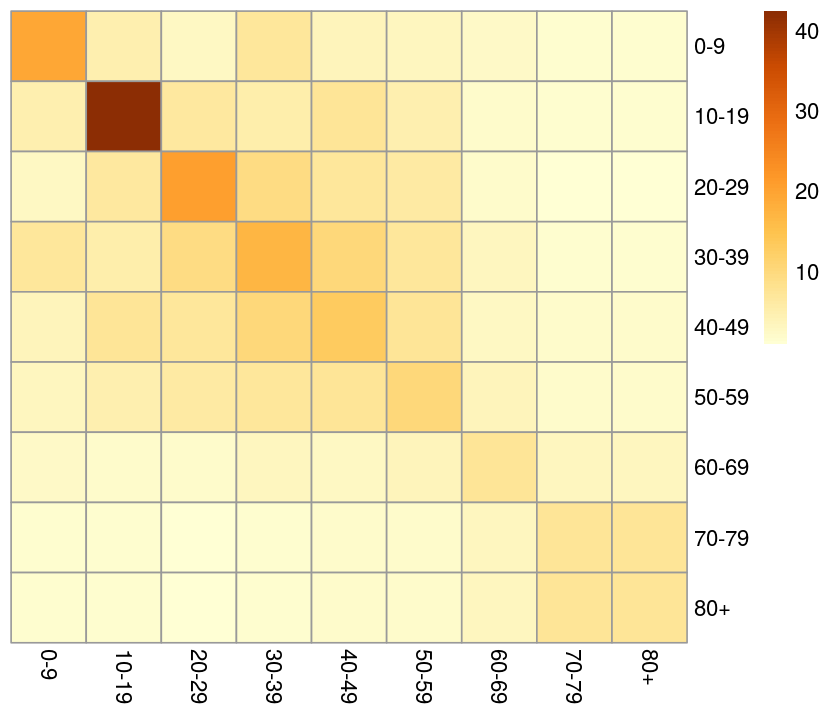}
    \caption{The contact matrix we use, derived from the contact matrix generated in \cite{Contact} for the United States.}
    \label{fig:my_label}
\end{figure}
\label{s.Contact}
We use the contact matrix generated for the United States in \cite{Contact}.  Note that matrices provided in \cite{Contact} are actually quite asymmetric, because they correspond to frequencies of interactions, which are affected also by the proportions of the populations belonging to different age group.  (In particular, if all groups were equally likely to interact, \cite{Contact} would report a contact matrix in which each row was equal to the population-proportion vector for the age groups).   The dynamics we describe in \eqref{dIdt}, etc., already account for the proportion of the susceptible population which belongs to each age group, thus we process the contact matrix by dividing column by the proportion of the population in the corresponding age group.  This results in a nearly symmetric matrix, which we symmetrize by taking the average of pairs of elements reflected across the diagonal.  We also bin the 5-year age groups used in \cite{Contact} to correspond to our 10-year age groups, and treat the 80+ and 70-79 groups identically for the purposes of the contact matrix, since \cite{Contact} does not provide separate contact data for the 80+ age group.

Note that one could use location-specific contact matrices (for work interactions, home interactions, \emph{etc}), but we have eschewed this choice at this stage since doing so requires a model of how these different types of interactions coordinate to contribute to empirically observed $R_0$ values.

\section{Results}
Our simulations all begin from an initial infection affecting $100,000$ individuals (distributed proportional to the early-stable proportion vector) in an otherwise fully susceptible population of size roughly $3\times 10^8$.  In all of our scenarios, we assume that normal transmission levels are linearly resumed between the 9- and 15-month marks from the start of the simulation.

\begin{figure}[p]
\centering
\includegraphics[width=.77\linewidth]{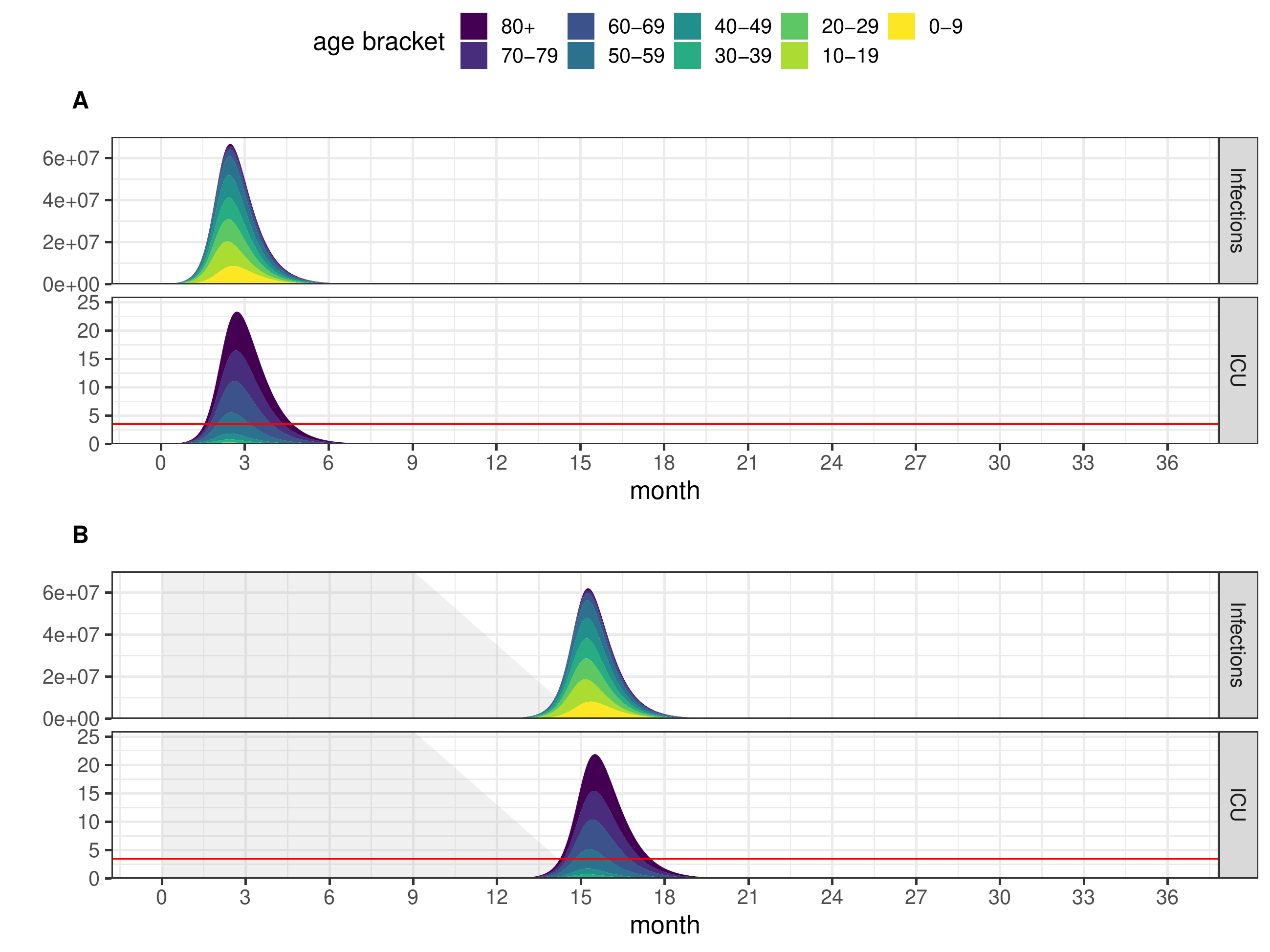}
\vspace{-1em}
\caption{\label{fig:no}\textbf{A}: No mitigations.  (TM: $15\cdot 10^5$).  \textbf{B}: Containment followed by resumption in transmission levels between 9 and 15 months (TM: $15\cdot 10^5$). }
\end{figure}

As a first example, Figure \ref{fig:no}\textbf{A} models the epidemic in the absence of any mitigations at all.  Figure \ref{fig:no}\textbf{B} models the epidemic in the presence of strict mitigations which, as in all our scenarios, are gradually relaxed between the 9- and 12-month marks.  Each figure shows the size of the infected population over time, and the ICU utilization over time, by age group.  Each figure caption indicates Total Mortalities (TM) from the scenario.  Note the light gray shading in the figures modeling mitigations serve as a reminder that normal transmission is linearly resumed between the 9- and 15- month marks.

\begin{figure}[p]
    \centering
    \includegraphics[width=.77\linewidth]{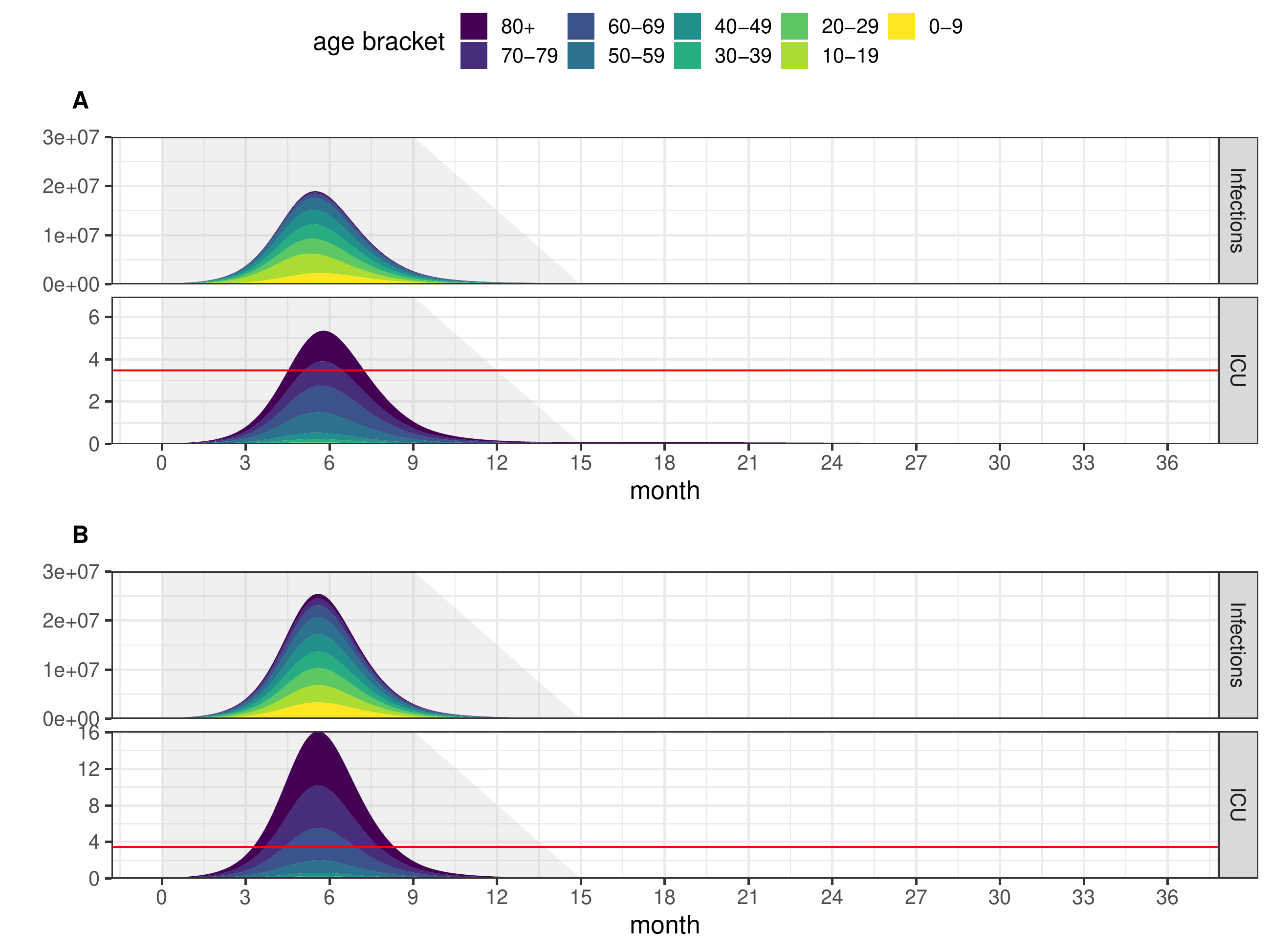}
\vspace{-1em}
\caption{\label{fig:homogeneous} \textbf{A}: Optimum homogeneous mitigations. TM: $7.8\cdot 10^5.$ \textbf{B}: Homogeneous measures without the contact matrix.  TM: $19.14\cdot 10^5.$}
\end{figure}

\subsection{The power of mitigations and natural heterogeneity}
In all our scenarios, we assume that transmission rates linearly resume between the 9 month and 15 month points.

Figure \ref{fig:homogeneous}\textbf{A} shows the outcome of optimal homogeneous measures.  That is, among all transmission reductions which could be applied equally to the all age groups and then gradually resumed as we assume, this is the choice of transmission reduction which minimizes deaths.  Transmission levels in this scenario are reduced by 40\% for all age groups in this scenario.  Note that this results in a reduction of mortalities by nearly 50\%.

Figure \ref{fig:homogeneous}\textbf{B} shows the outcome of the same mitigations ignoring the role of natural contact patterns in the population.  In particular, for this scenario we have assumed a counterfactual where the likelihood of two people of different ages interacting is determined just by the relative sizes of the populations of the age-groups; this corresponds to a constant $\ccc$ matrix.  We see 2.5 times more mortalities in this scenario, demonstrating the power of natural population heterogeneity to reduce infection mortality for COVID-19. (The optimum choice for transmission is roughly the same for the uniform contact matrix.)

In all our figures, the scale ICU utilization is shown rescaled by a factor of $\frac{10,000}{\textrm{total population}}$.  In other words, units for the ICU utilization figures is the ICU capacity---per 10,000 people---required to support ICU admissions at that level.  The red line shows a nominal capacity level of 3.47 beds per 10,000 people.
\subsection{Age-targeted mitigations}
\label{s.scenarios}

For our age-targeted mitigations, we consider  relaxing mitigations just on those under 40, just on those under 50, and just on those under 60.

Since it is natural to expect targeted mitigations to be  based on household-level of risks, because of cohabitation of younger and older adults, we consider, in each case, a scenario where only 2/3 of the younger population is subject to normal transmission levels.

\begin{figure}[p]
    \centering
    \includegraphics[width=.77\linewidth]{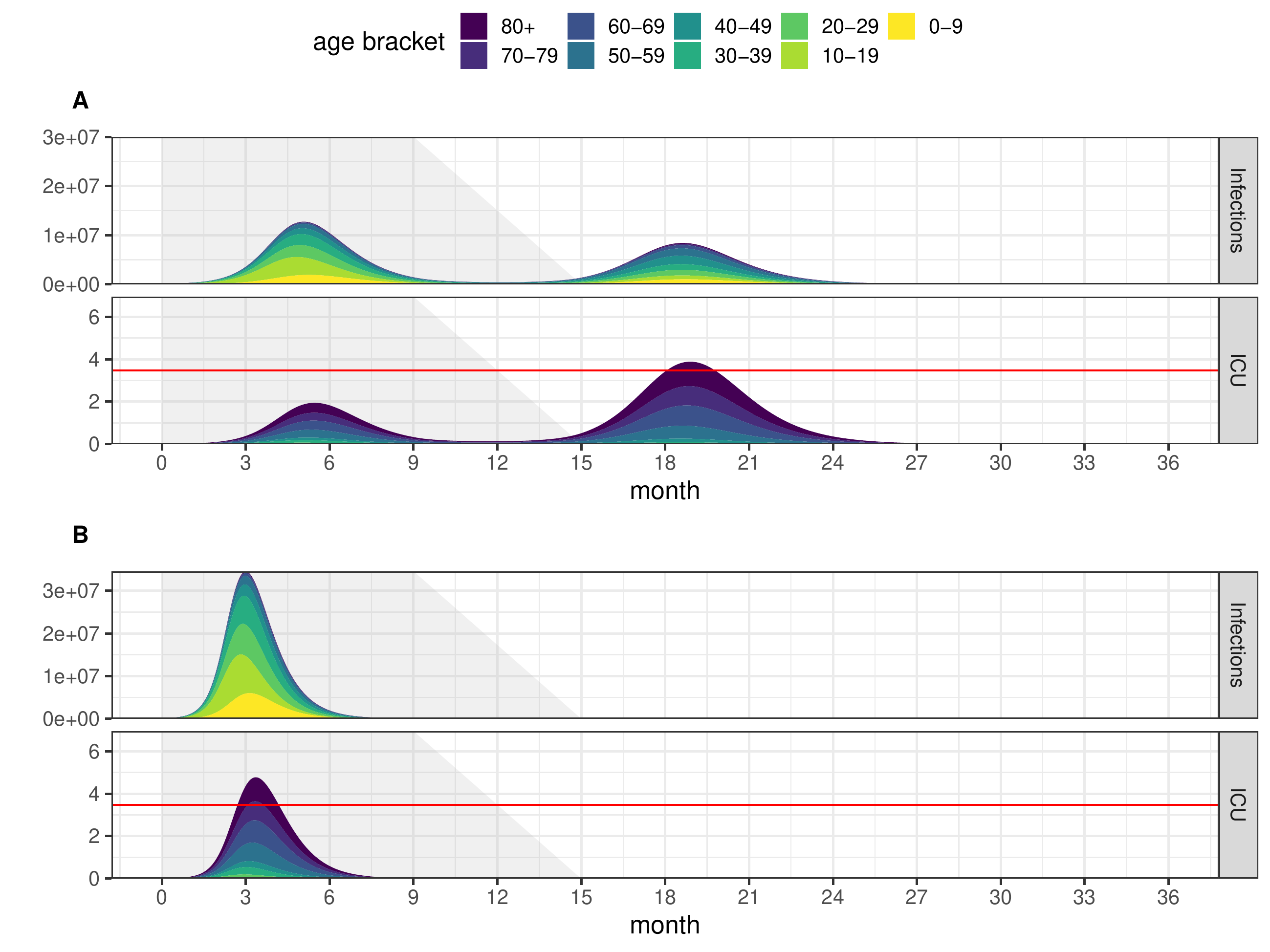}
    \vspace{-1em}
    \caption{\textbf{A}: 2/3 of people of under 39 released at normal levels %, rest depressed by 70\% 
    (TM: $9.0\cdot 10^5$).  \textbf{B}: All of people of under 39 released at normal levels %, rest depressed by 70\% 
    (TM: $3.9\cdot 10^5$).}
    \label{fig:39}
\end{figure}

\begin{figure}[p]
    \centering
    \includegraphics[width=.77\linewidth]{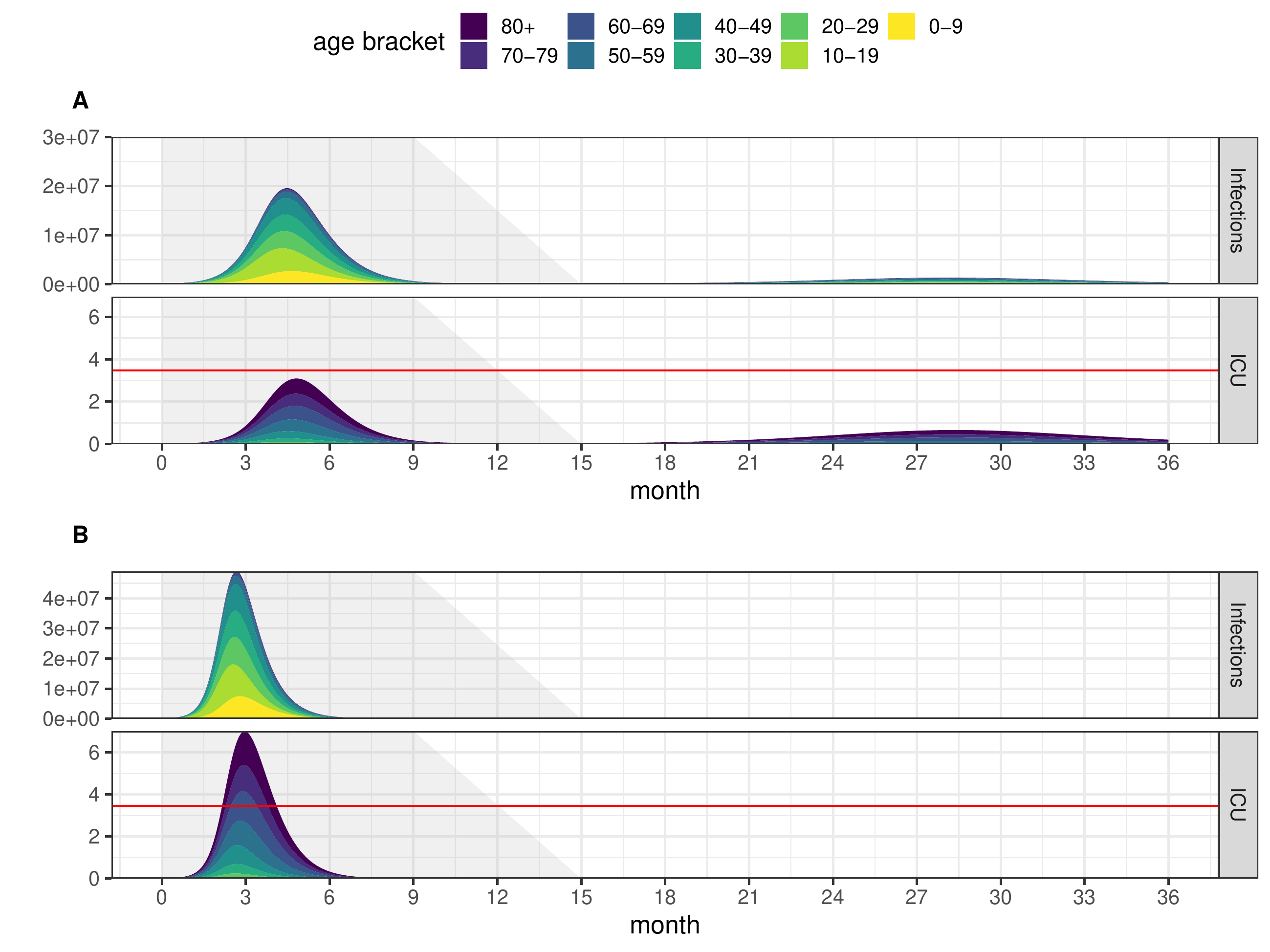}
    \vspace{-1em}
    \caption{\textbf{A}: 2/3 under 49 released at normal levels %, rest depressed by 70\% 
    (TM: $6.0\cdot 10^5$).  
    \textbf{B}: All under 49 released at normal levels %, rest depressed by 70\% 
    (TM: $5.0\cdot 10^5$).}
    \label{fig:49}
\end{figure}

\begin{figure}[p]
    \centering
    \includegraphics[width=.77\linewidth]{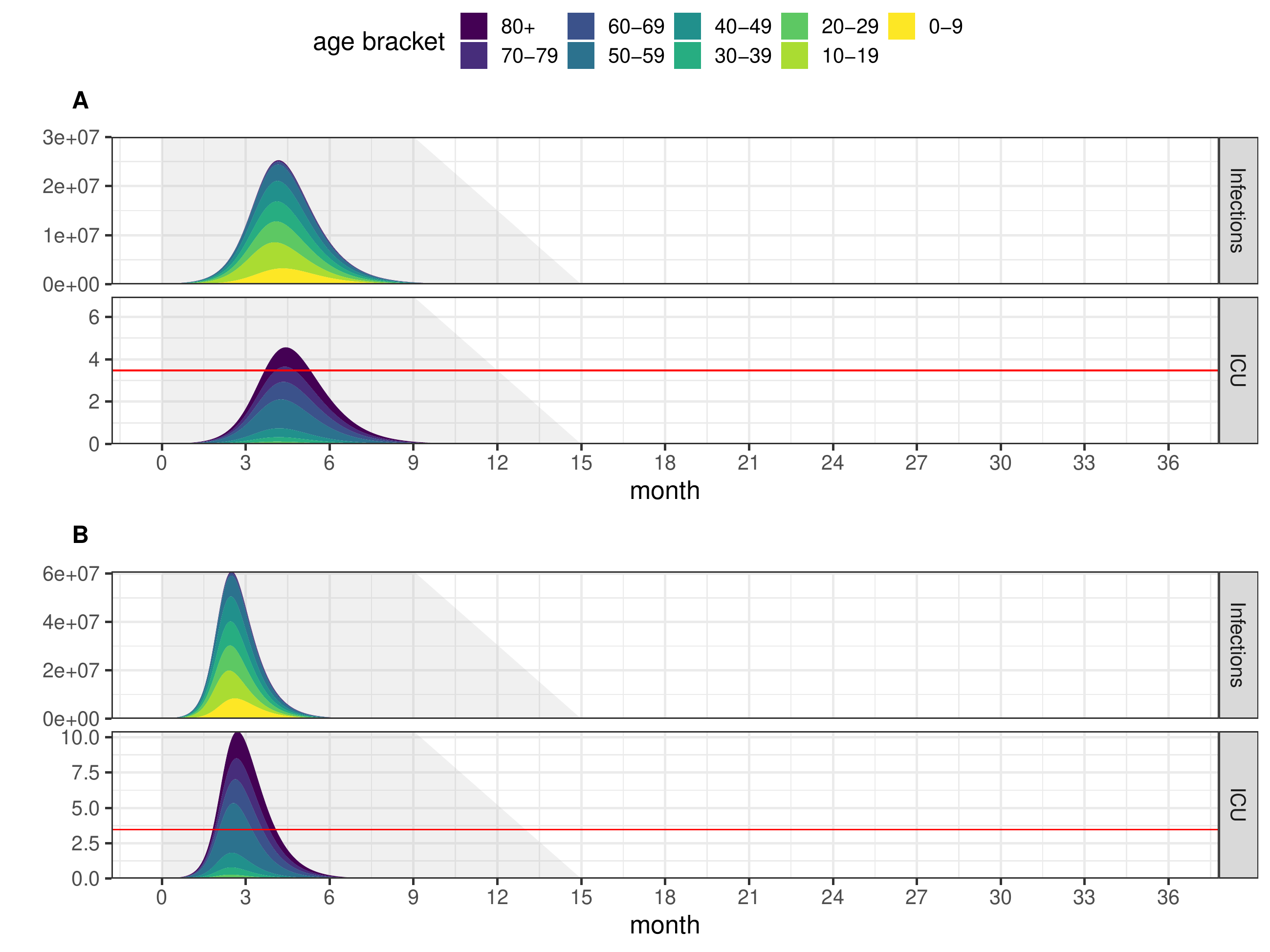}
    \vspace{-1em}
    \caption{\textbf{A}: 2/3 under 59 released at normal levels %, rest depressed by 70\% 
    (TM: $4.53\cdot 10^5$).  
    \textbf{B}: All under 59 released at normal levels %, rest depressed by 70\% 
    (TM: $6.66\cdot 10^5$).}
    \label{fig:59}
\end{figure}

In each of these scenarios, depicted in Figures \ref{fig:39}, \ref{fig:49}, and \ref{fig:59}, we assume that the relaxed population is subject to normal transmission levels, while transmission to, from, and within the rest of the population is depressed by $70\%$ from normal levels.  (Results for other choices for these and other constants are discussed in Section \ref{s.sensitivity}.)

What we see in Figures \ref{fig:39}, \ref{fig:49}, and \ref{fig:59} is that age-targeting has the potential to greatly reduce total mortalities compared with optimum choices for homogeneous measures.  At the same time, we see that the best strategies for age-targeting are sensitive to the fraction of the younger population which can be released.   In general, if too few people are released initially, a second wave occurs when transmission levels return to normal.  Conversely, if too many are released, ICU utilization is high in the first wave.  Thus the optimum choice for the age cutoff depends on the fraction of people we expect below the age cutoff to actually be released.

\section{Sensitivity of analysis}
\label{s.sensitivity}
In Section \ref{s.scenarios} we presented scenarios of age-targeted mitigations where $R_0=2.7$ and transmission levels for the groups subject to mitigations are reduced by 70\%.  In Figures \ref{fig:2.4_0.1}, \ref{fig:2.4_0.3}, \ref{fig:2.4_0.5}, \ref{fig:2.7_0.1}, \ref{fig:2.7_0.3}, \ref{fig:2.7_0.5}, \ref{fig:3_0.1}, \ref{fig:3_0.3}, \ref{fig:3_0.5} we show scenarios where $R_0$ is $2.4$, $2.7$, or $3.0$, and the transmission reduction for the group subject to mitigations is $90\%$, $70\%$, or $50\%$.  (We refer to these as very strict mitigations, strict mitigations, and moderate mitigations.)  Note that the panel for $R_0=2.7$ with strict mitigations ($70\%$ reduction) includes the scenarios discussed in Section \ref{s.scenarios}.

In each of these figure panels, scenarios where peak ICU utilization does not exceed the nominal ICU capacity by more than 50\% are highlighted.  Note that this includes some scenarios which exhibit a large second wave and thus exhibit many mortalities.

The figures for these scenarios all plot 36 months of simulation.  In a few cases, a small but non-negligible part of the second wave occurs past this point.  Mortality and ICU utilization for each scenario is summarized in Table \ref{tab:scenarios}; this data is based on a 10 year simulation period, to capture second waves extending beyond 36 months.

\section{Age-specific mortality impact}
\begin{figure}[t]
    \centering
    \includegraphics[width=\linewidth]{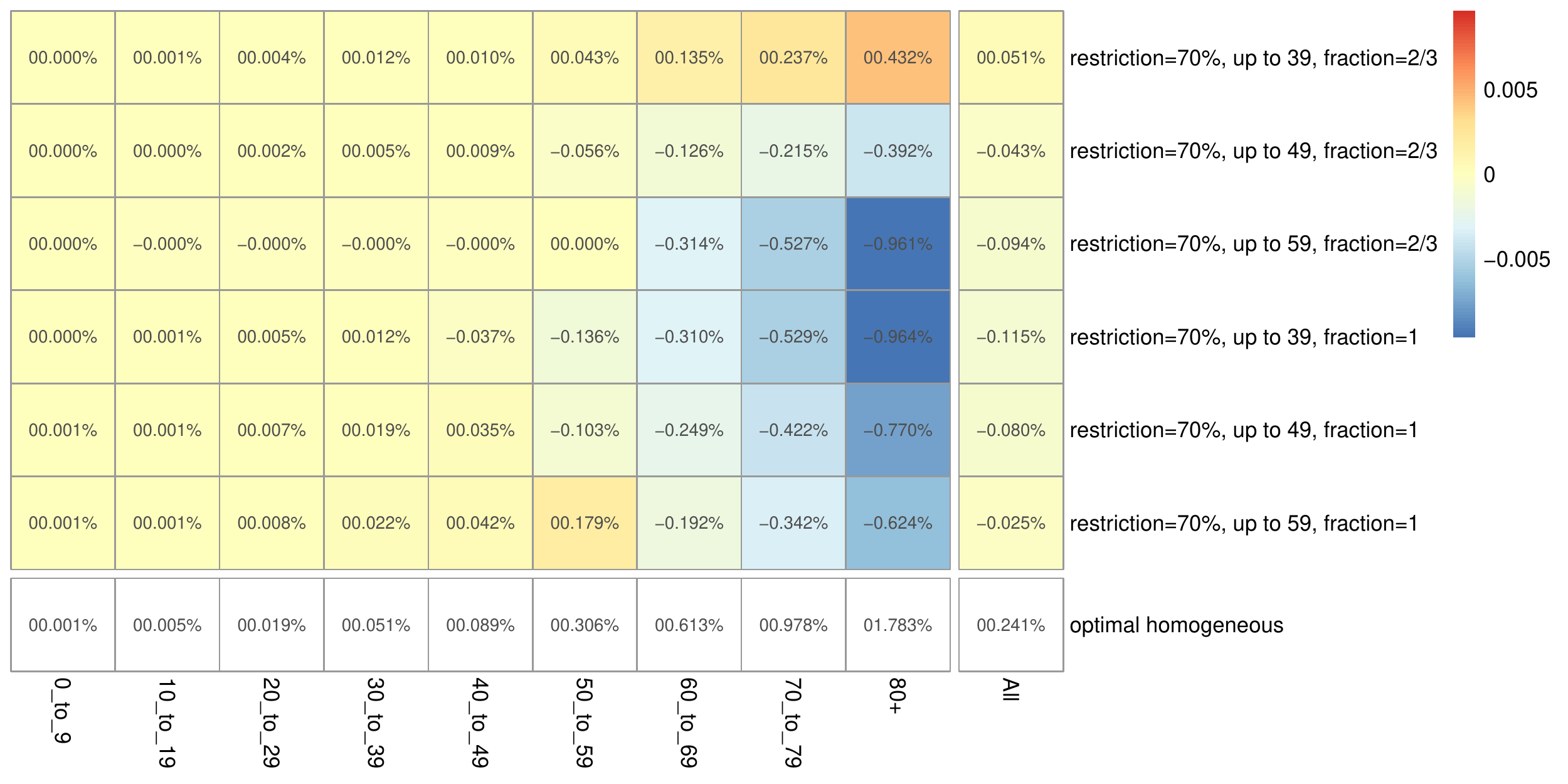}
    \caption{Impact on mortalities by age group.  The bottom row of this table shows mortalities in each age group for the optimum homogeneous mitigations scenario, as a fraction of the total population of the age group.  Each other row corresponds to a scenario discussed in Section \ref{s.scenarios}. The values in the cells in these rows show the \emph{difference} in the mortality rate for each age group for the given scenario, compared with the optimum homogeneous scenario.}
    \label{fig:age-mort}
\end{figure}

\begin{figure}[p]
    \centering
    \includegraphics[width=.93\linewidth]{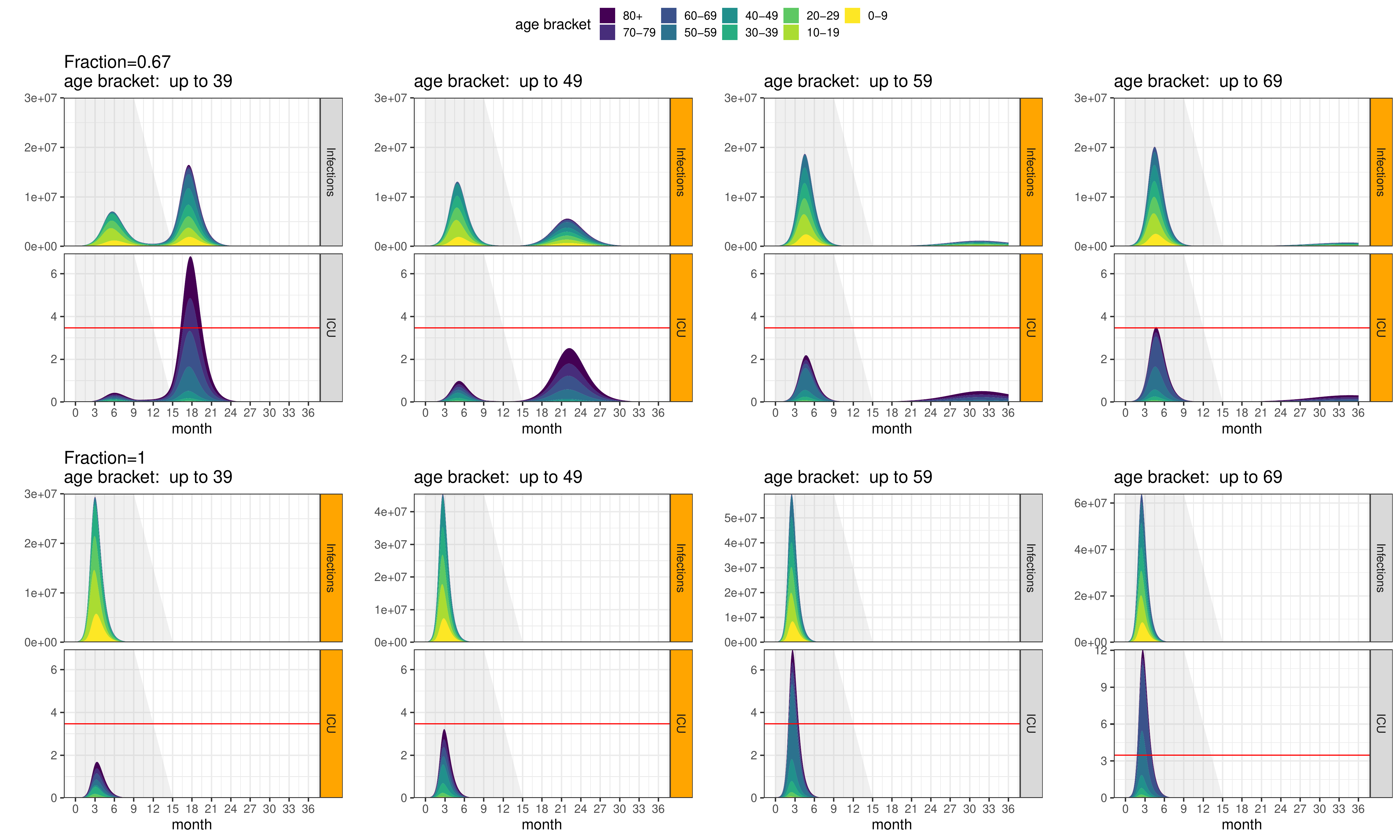}
    \vspace{-1em} 
    \caption{Scenarios for $R_0=2.4$, very strict mitigations (90\% reduction in transmission for groups subject to mitigations).  Scenarios which do not exceed the nominal ICU capacity by more than 50\% are highlighted.}
    \label{fig:2.4_0.1}
\end{figure}

\begin{figure}[p]
    \centering
    \includegraphics[width=.93\linewidth]{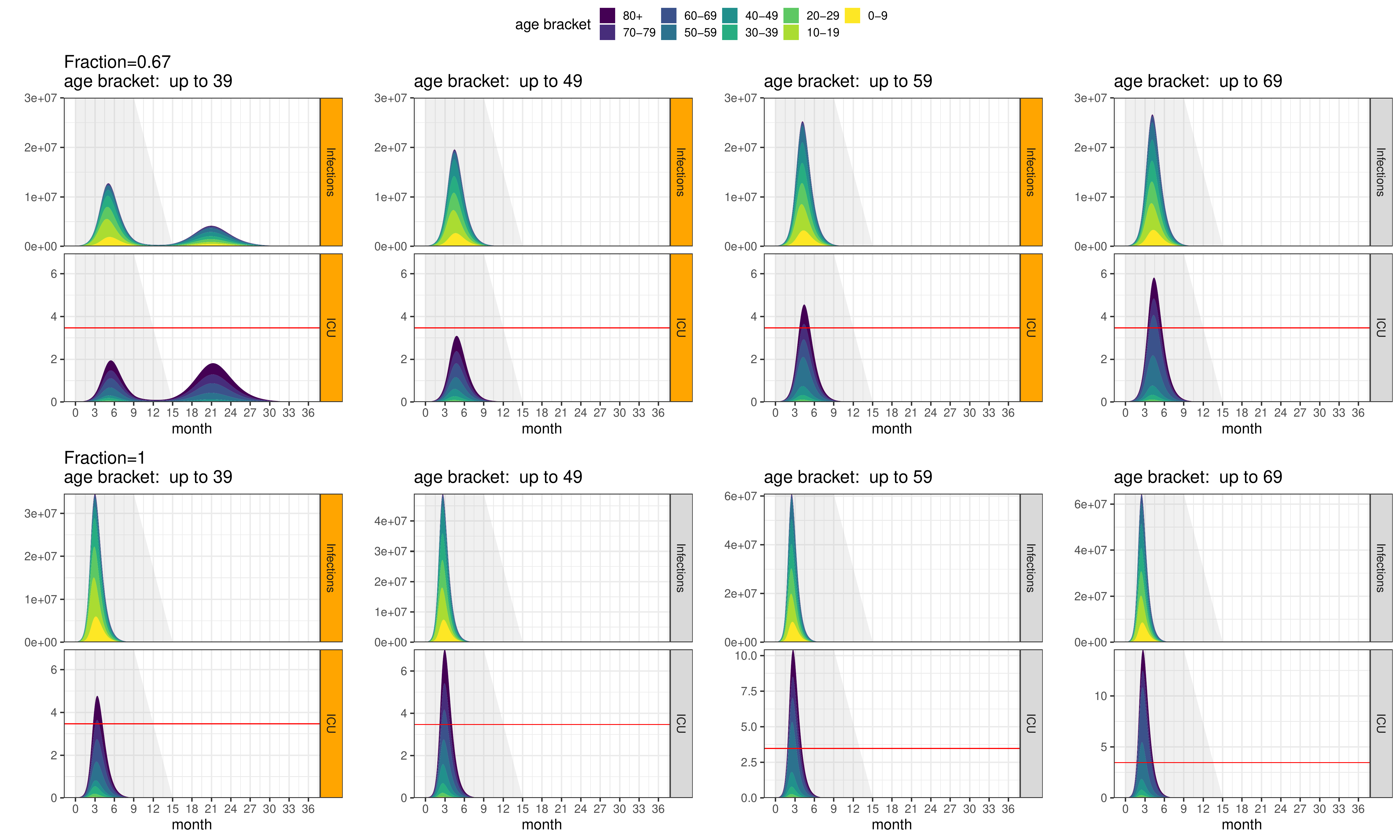}
    \vspace{-1em} 
    \caption{Scenarios for $R_0=2.4$, strict mitigations (70\% reduction in transmission for groups subject to mitigations).  Scenarios which do not exceed the nominal ICU capacity by more than 50\% are highlighted.}
    \label{fig:2.4_0.3}
\end{figure}

It might be natural to suspect that age-specific strategies are simply trading mortalities in one group for mortalities in another.  However, we find in our models that age-targeted restrictions can dramatically reduce mortalities among older populations with very small impacts on mortality in younger populations; see Figure \ref{fig:age-mort}.  In particular, because of the effect age-targeted measures can have on ICU over-utilization (whose impact on mortality we have not modeled), it is quite possible that well-calibrated age-targeted mitigations could improve typical outcomes for all age groups.

\section{Discussion and caveats}
\begin{figure}[p]
    \centering
    \includegraphics[width=.93\linewidth]{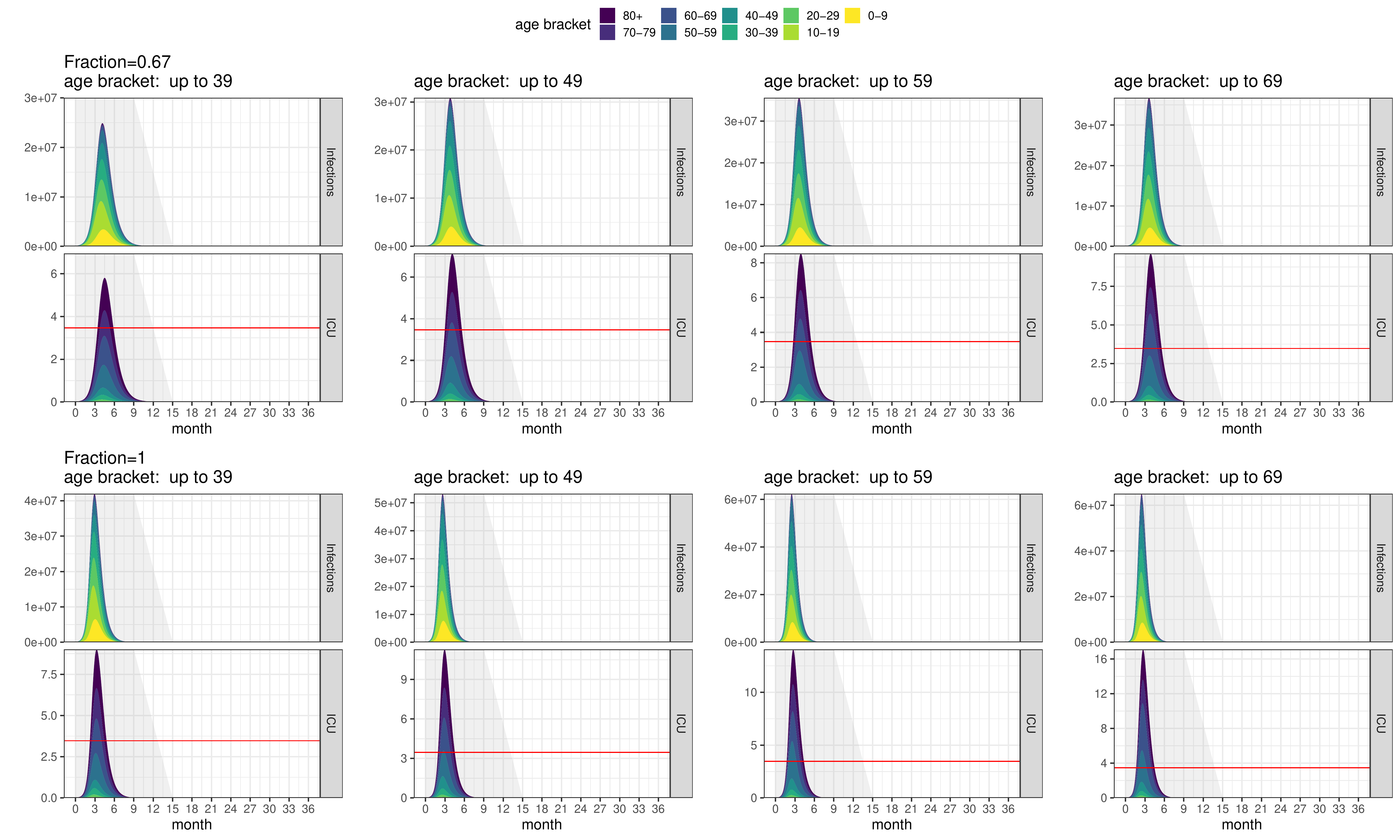}
    \vspace{-1em} 
    \caption{Scenarios for $R_0=2.4$, moderate mitigations (50\% reduction in transmission for groups subject to mitigations).  All of these scenarios exceed the nominal ICU capacity by more than 50\%.}
    \label{fig:2.4_0.5}
\end{figure}

\begin{figure}[p]
    \centering
    \includegraphics[width=.93\linewidth]{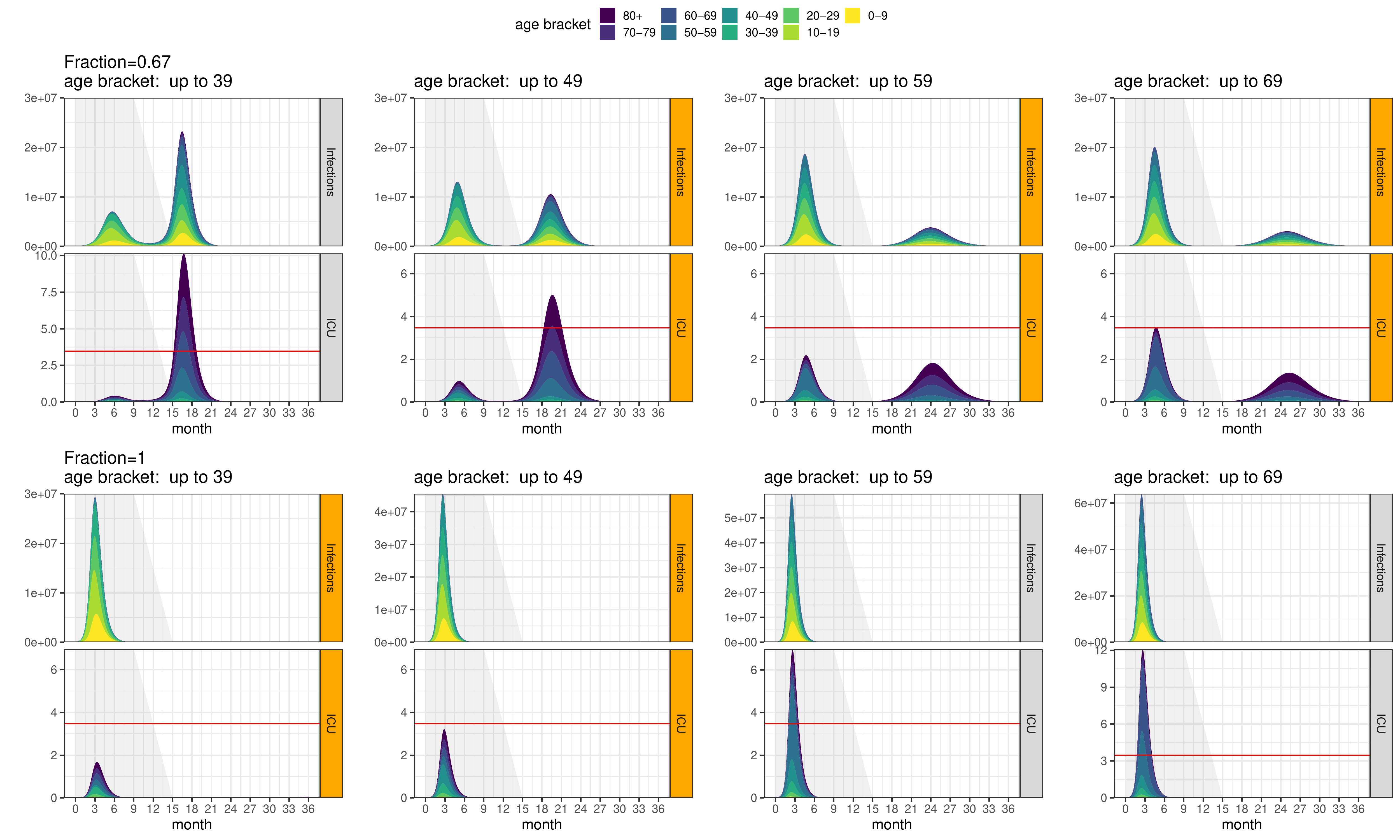}
    \vspace{-1em} 
    \caption{Scenarios for $R_0=2.7$, very strict mitigations (90\% reduction in transmission for groups subject to mitigations).  Scenarios which do not exceed the nominal ICU capacity by more than 50\% are highlighted.}
    \label{fig:2.7_0.1}
\end{figure}

\begin{figure}[p]
    \centering
    \includegraphics[width=.93\linewidth]{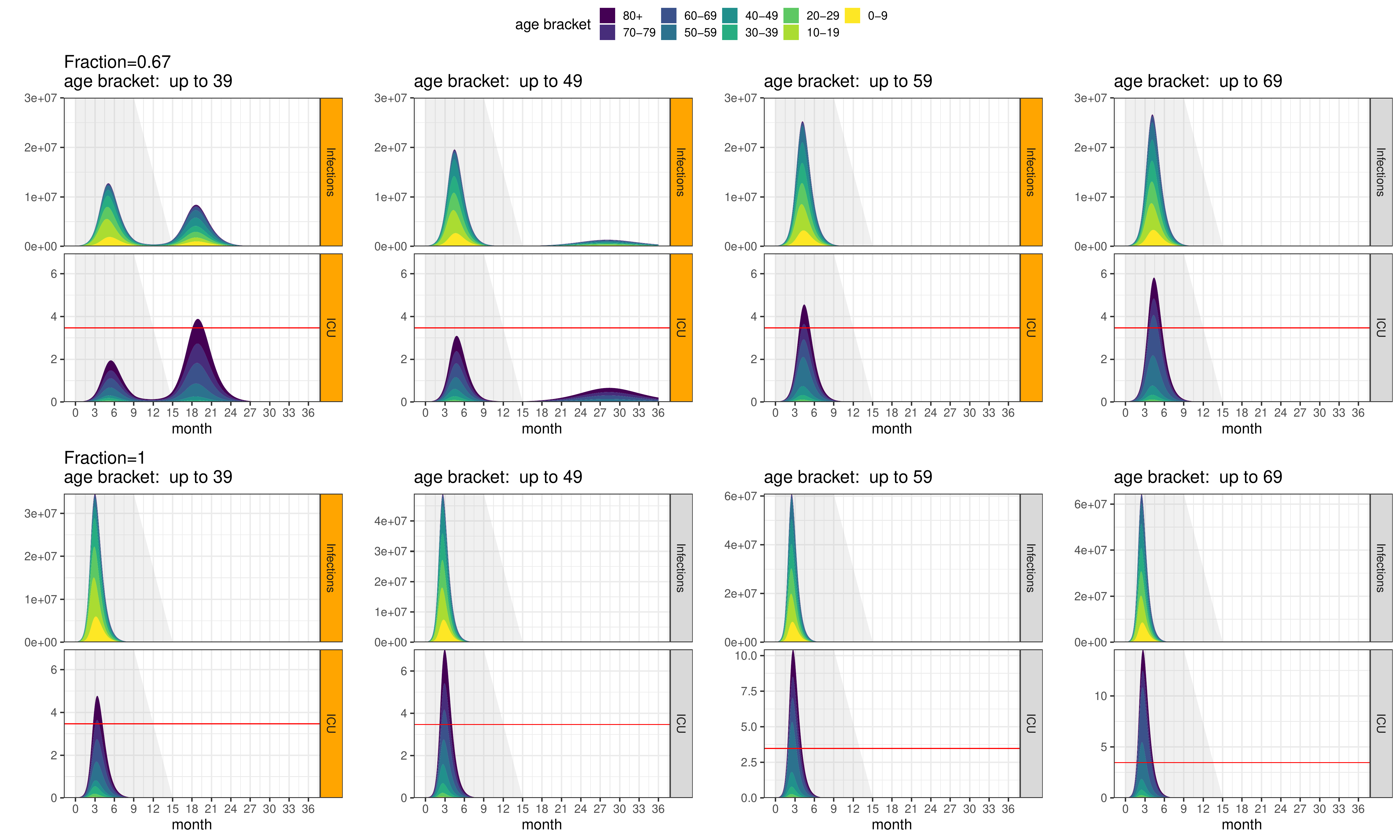}
    \vspace{-1em} 
    \caption{Scenarios for $R_0=2.7$, strict mitigations (70\% reduction in transmission for groups subject to mitigations).  Scenarios which do not exceed the nominal ICU capacity by more than 50\% are highlighted.}
    \label{fig:2.7_0.3}
\end{figure}

\begin{figure}[p]
    \centering
    \includegraphics[width=.93\linewidth]{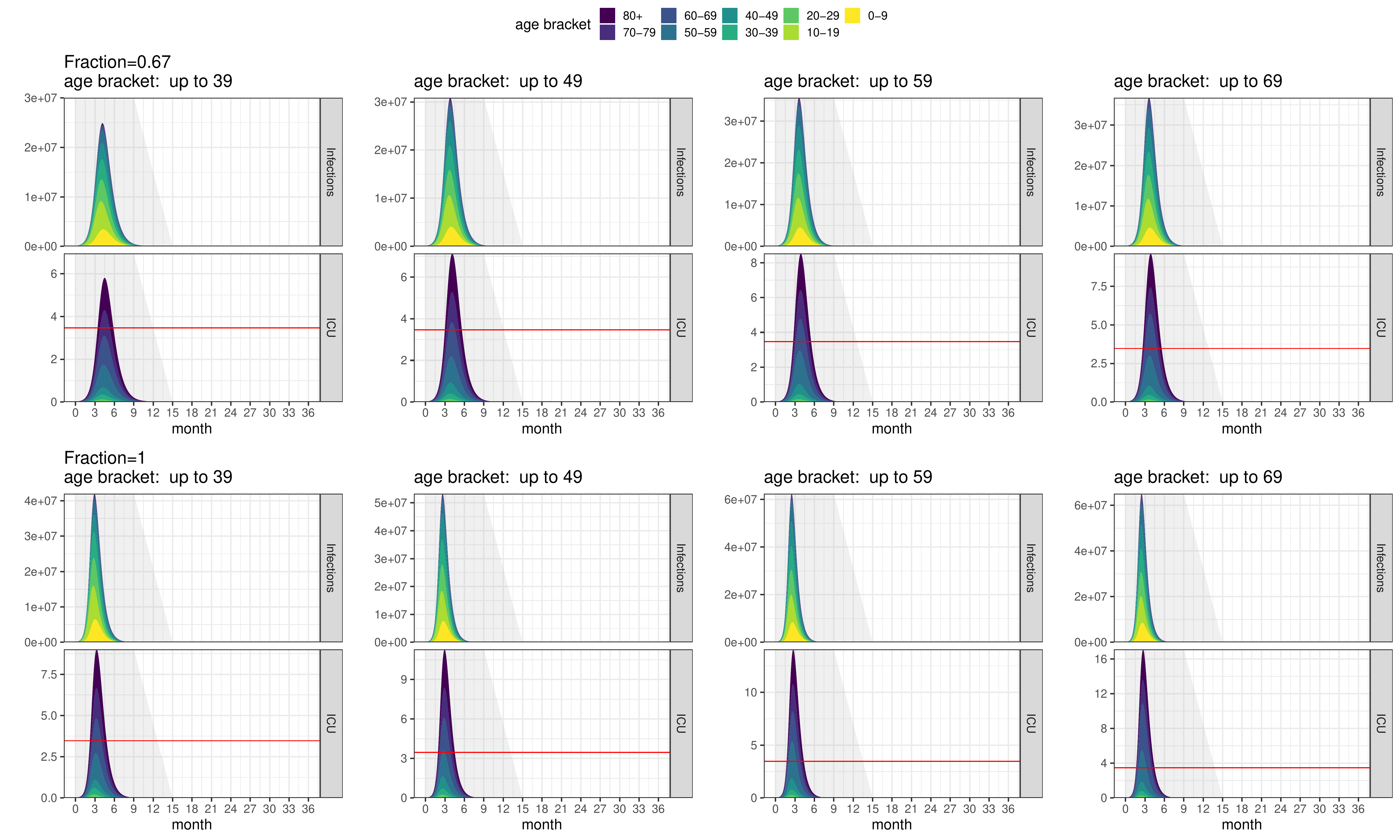}
    \vspace{-1em} 
    \caption{Scenarios for $R_0=2.7$, moderate mitigations (50\% reduction in transmission for groups subject to mitigations).  All of these scenarios exceed the nominal ICU capacity by more than 50\%.}
    \label{fig:2.7_0.5}
\end{figure}

\begin{figure}[p]
    \centering
    \includegraphics[width=.93\linewidth]{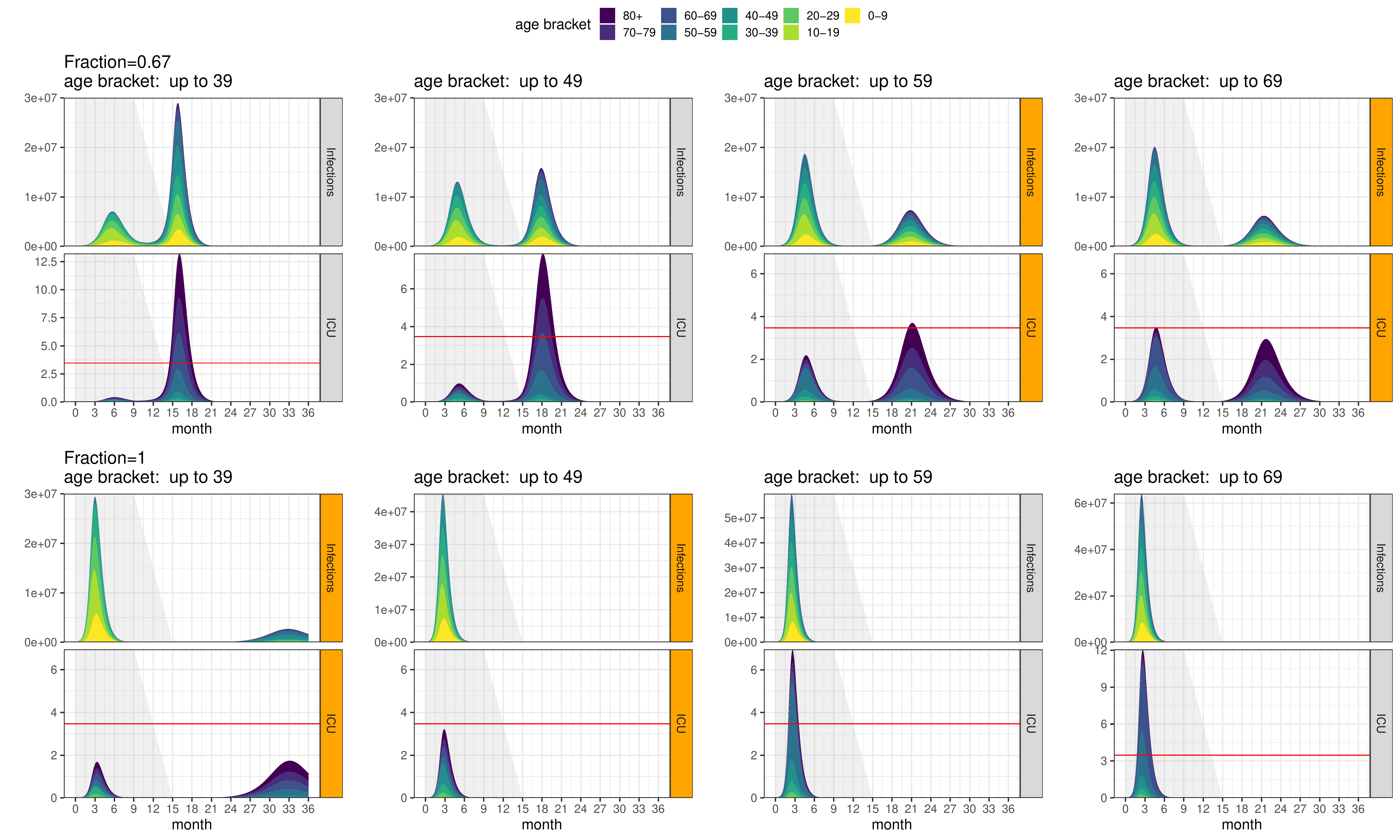}
    \vspace{-1em} 
    \caption{Scenarios for $R_0=3.0$, very strict mitigations (90\% reduction in transmission for groups subject to mitigations).  Scenarios which do not exceed the nominal ICU capacity by more than 50\% are highlighted.}
    \label{fig:3_0.1}
\end{figure}

\begin{figure}[p]
    \centering
    \includegraphics[width=.93\linewidth]{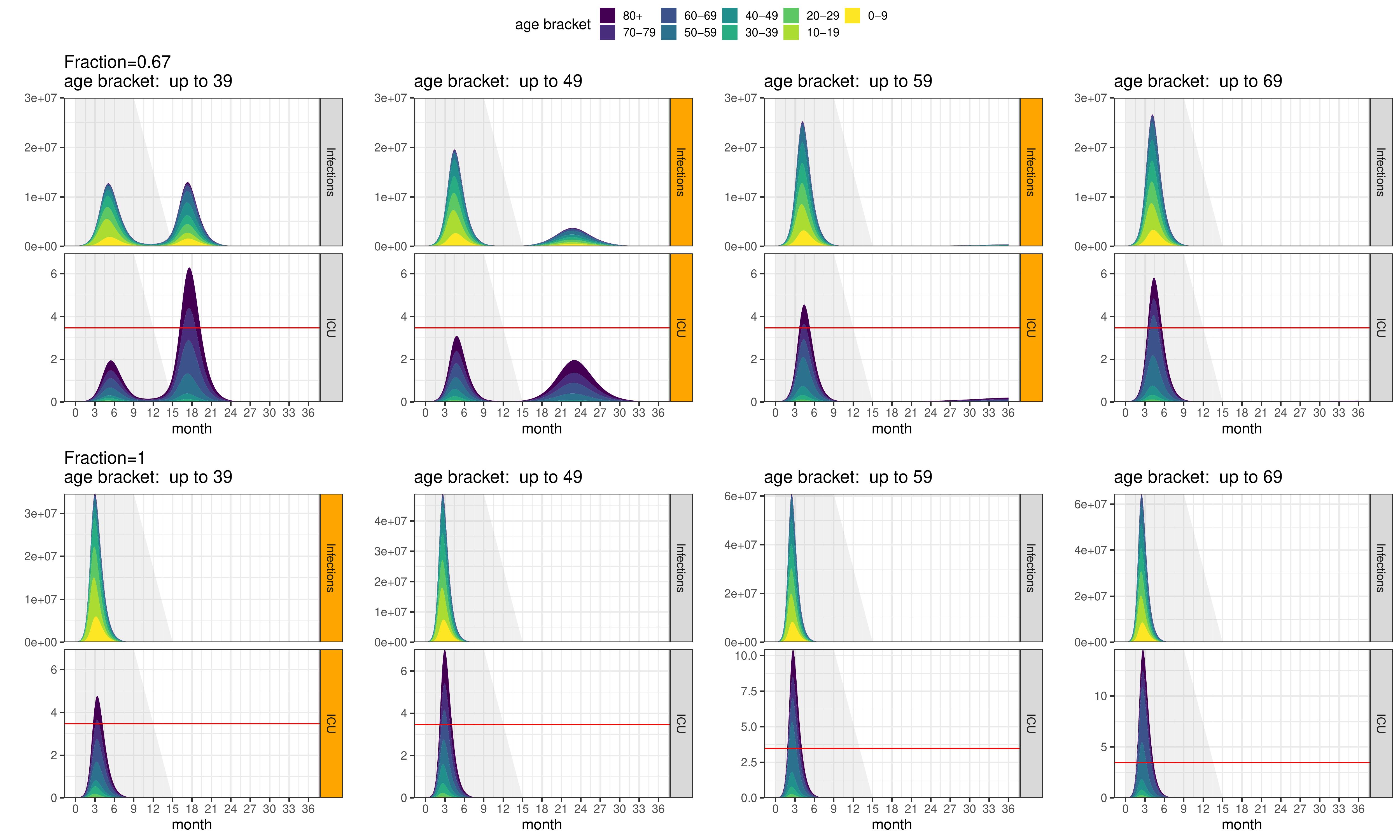}
    \vspace{-1em} 
    \caption{Scenarios for $R_0=3.0$, strict mitigations (70\% reduction in transmission for groups subject to mitigations).  Scenarios which do not exceed the nominal ICU capacity by more than 50\% are highlighted.}
    \label{fig:3_0.3}
\end{figure}

\begin{figure}[p]
    \centering
    \includegraphics[width=.93\linewidth]{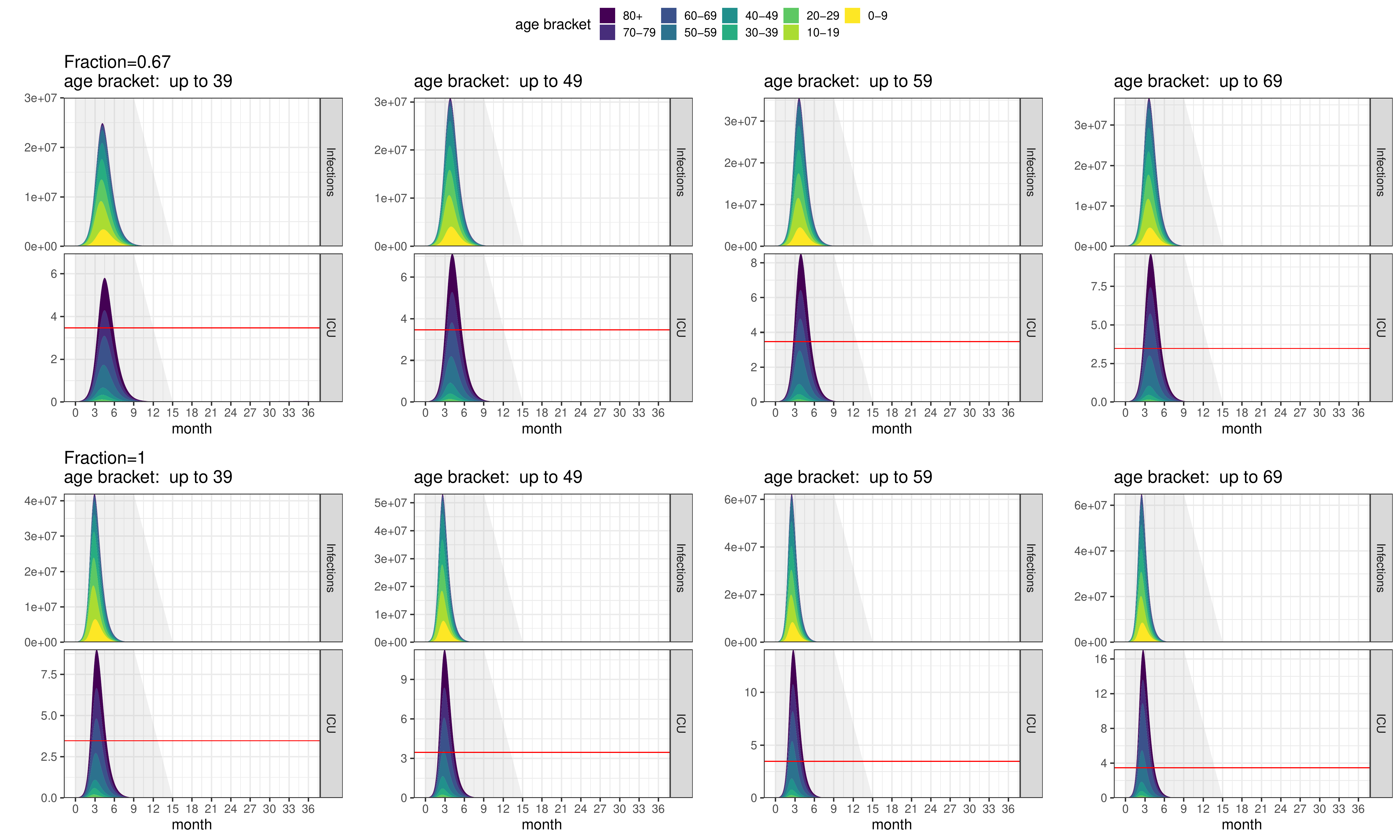}
    \vspace{-1em} 
    \caption{Scenarios for $R_0=3.0$, moderate mitigations (50\% reduction in transmission for groups subject to mitigations).  All of these scenarios exceed the nominal ICU capacity by more than 50\%.}
    \label{fig:3_0.5}
\end{figure}

We have considered a model of age-heterogeneous transmission and mitigation in a COVID-19-like epidemic, which is simple but also tied to current estimates of both disease parameters and U.S.-specific contact patterns.  We find that age-targeted mitigations can have a dramatic effect both on mortality and ICU utilization.  However, we also find that to be successful, age-targeted mitigations may have to be strict.  Our scenarios modeling moderate mitigations on the restricted group (shown in Figures \ref{fig:2.4_0.5}, \ref{fig:2.7_0.5}, and \ref{fig:3_0.5}) fare quite poorly (see also Table \ref{tab:scenarios}).

Importantly, we find that while relatively good strategies exist in a range of scenarios, so long as mitigations on restricted groups can be strict or very strict, the precise choices which minimize ICU utilization and deaths are sensitive---for example, to the fraction of younger people which will actually be released from mitigations. 

We also find that if only moderate mitigations are possible on the population subject to mitigations, then the discrete set of age-targeted mitigations we considered fared poorly.

We view our modeling as demonstrating a qualitative point: strict age-targeted mitigations can have a powerful effect on mortality and ICU utilization.  We expect that public policy motivated by this kind of finding would have to be responsive; for example, by relaxing restrictions on larger and large groups conservatively, while monitoring the progress of the epidemic.

Note in Table \ref{tab:scenarios}, for every $R_0$ value, there are heterogeneous strategies that outperform the optimal homogeneous strategies.  There are also heterogeneous strategies that do worse, sometimes much worse.  These poor-performing strategies fall into one of two types:
\begin{itemize}
    \item Those with too large a relaxed population, so that the initial epidemic is not sufficiently constrained, and
    \item Those with too small a relaxed population, so that when transmission rates resume, a second wave results (which disproportionately effects the older population).
    \end{itemize}
    The simple lessons for policy are twofold: first, \textbf{initial relaxations must be responsive to conditions on the ground}, e.g., through monitoring of ICU utilization.  Second, \textbf{the return to normal must also proceed cautiously} and, ideally, be informed by data on the size of the infected population.  Although we have considered simple 2-stage strategies here, it would be natural to consider implementing policies with more stages, as a way of proceeding cautiously.
\bigskip

Any predictive model is an oversimplification of the real world whose predictions depend on parameter values whose true values will only be known after-the-fact. The model we employ is particularly simple, and while this simplicity can be an asset when demonstrating qualitative phenomena, it also presents obvious limitations. For example:
 \begin{itemize}
    \item We do not model seasonality, since the effects of seasonality on COVID-19 remain unsettled. Seasonal forcing could mean, for example, that after relaxing restrictions on some group, they may have to be reinstated as transmission rates increase.
    \item While we do use (simple) models of known age-group contact patterns, we don't model the effects of specific mitigations on those patterns; for example, we don't evaluate the specific effects of things like closing schools. We also don't attempt to model the different effects of mitigations on within-home and out-of-home contact patterns, since the ways these each contribute to empirically observed $R_0$ values is complicated.

    \item There is still considerable uncertainty regarding basic parameters of COVID-19 such as its transmissibility, the infection mortality rate, and the ICU admission rate. While we have relied on expert choices for the parameters we have used, all quantitative findings we make (such as ICU admissions) are sensitive to these choices.
\end{itemize}

On the other hand, there are ways in which our analysis has been conservative. For example:
\begin{itemize}
    \item When modeling homogeneous mitigation strategies in Figure 2, we allow ourselves complete freedom to choose an effective $R_0$ value resulting from mitigations. On the other hand, for our examples of heterogeneous mitigation strategies, we have tied our hands considerably more. We simply allow ourselves to choose which age group, in intervals of 10, to release to normal transmission levels---we have just a few discrete options to choose from. In particular, if were allowed ourselves to combine mild mitigations on the relaxed group with strong mitigations on the rest, we would be able to achieve fewer mortalities and ICU admissions.  Note that many of our age-targeted mitigation scenarios exhibit epidemics which end well before the 9-month mark; these curves have room to be flattened.

    \item We have considered scenarios where all the younger age group will be able to have relaxed transmission rates, because of cohabitation with older household members, or because of risk factors other than age. This can make our analysis worse, by making the average age of the required immunity herd. However, we have not taken advantage of a presumed benefit that this would confer: if effective risk-models incorporating factors beyond age could be deployed, it is plausible that the ICU admission rate for the relaxed groups could be decreased.

    \item We have not modeled flexibility in ICU capacity.
\end{itemize}

\noindent\textbf{Acknowledgements}
We have benefited from discussion and helpful comments with regard to this and our previous analysis \cite{covid.html} with many people, including Boris Bukh, Don Burke, Forrest Collman, Jordan Ellenberg, Ryan O'Donnell, Russell Schwartz, among several others.

\begin{table}[]
    \centering \tiny
\begin{tabular}{l|llllrr}
  \hline
 $R_0$ & strategy & supression & relaxed ages & fraction & mortalities & peak.ICU \\ 
  \hline\hline
  2.4 & homogeneous & optimal & NA & NA & 6.79 & 4.64 \\
    \hline
  2.4 & heterogeneous & 90\% & up to 39 & 0.67 & 9.26 & 6.82 \\ 
\rowcolor{highlight}  2.4 & heterogeneous & 90\% & up to 49 & 0.67 & 6.45 & 2.52 \\ 
\rowcolor{highlight}  2.4 & heterogeneous & 90\% & up to 59 & 0.67 & 4.51 & 2.19 \\ 
  \rowcolor{highlight}2.4 & heterogeneous & 90\% & up to 69 & 0.67 & 5.29 & 3.51 \\ 
\rowcolor{highlight}  2.4 & heterogeneous & 90\% & up to 39 & 1 & 1.35 & 1.69 \\ 
\rowcolor{highlight}  2.4 & heterogeneous & 90\% & up to 49 & 1 & 2.21 & 3.22 \\ 
\rowcolor{highlight}   2.4 & heterogeneous & 90\% & up to 59 & 1 & 4.23 & 6.95 \\ 
  2.4 & heterogeneous & 90\% & up to 69 & 1 & 7.46 & 12.05 \\ 
  \hline
  2.4 & heterogeneous & 70\% & up to 39 & 0.67 & 6.96 & 1.95 \\ 
\rowcolor{highlight}  2.4 & heterogeneous & 70\% & up to 49 & 0.67 & 3.94 & 3.09 \\ 
\rowcolor{highlight}  2.4 & heterogeneous & 70\% & up to 59 & 0.67 & 4.51 & 4.56 \\ 
\rowcolor{highlight}  2.4 & heterogeneous & 70\% & up to 69 & 0.67 & 5.66 & 5.82 \\ 
\rowcolor{highlight}  2.4 & heterogeneous & 70\% & up to 39 & 1 & 3.89 & 4.78 \\ 
\rowcolor{highlight}  2.4 & heterogeneous & 70\% & up to 49 & 1 & 4.96 & 7.01 \\ 
\rowcolor{highlight}  2.4 & heterogeneous & 70\% & up to 59 & 1 & 6.66 & 10.42 \\ 
  2.4 & heterogeneous & 70\% & up to 69 & 1 & 9.20 & 14.55 \\ 
  \hline
\rowcolor{highlight}  2.4 & heterogeneous & 50\% & up to 39 & 0.67 & 6.09 & 5.81 \\ 
\rowcolor{highlight}  2.4 & heterogeneous & 50\% & up to 49 & 0.67 & 6.70 & 7.13 \\ 
  2.4 & heterogeneous & 50\% & up to 59 & 0.67 & 7.48 & 8.52 \\ 
  2.4 & heterogeneous & 50\% & up to 69 & 0.67 & 8.35 & 9.62 \\ 
  2.4 & heterogeneous & 50\% & up to 39 & 1 & 7.09 & 9.00 \\ 
  2.4 & heterogeneous & 50\% & up to 49 & 1 & 7.92 & 11.28 \\ 
  2.4 & heterogeneous & 50\% & up to 59 & 1 & 9.12 & 14.05 \\ 
  2.4 & heterogeneous & 50\% & up to 69 & 1 & 10.92 & 17.07 \\ 
  \hline
2.7 & homogeneous & optimal & NA & NA & 7.43 & 5.35 \\ 
\hline
  2.7 & heterogeneous & 90\% & up to 39 & 0.67 & 11.19 & 10.13 \\ 
  2.7 & heterogeneous & 90\% & up to 49 & 0.67 & 8.62 & 5.01 \\ 
\rowcolor{highlight}  2.7 & heterogeneous & 90\% & up to 59 & 0.67 & 6.57 & 2.19 \\ 
\rowcolor{highlight}  2.7 & heterogeneous & 90\% & up to 69 & 0.67 & 7.15 & 3.51 \\ 
\rowcolor{highlight}  2.7 & heterogeneous & 90\% & up to 39 & 1 & 2.27 & 1.69 \\ 
\rowcolor{highlight}  2.7 & heterogeneous & 90\% & up to 49 & 1 & 2.21 & 3.22 \\ 
\rowcolor{highlight}  2.7 & heterogeneous & 90\% & up to 59 & 1 & 4.23 & 6.95 \\ 
  2.7 & heterogeneous & 90\% & up to 69 & 1 & 7.46 & 12.05 \\ 
  \hline
  2.7 & heterogeneous & 70\% & up to 39 & 0.67 & 9.00 & 3.89 \\ 
\rowcolor{highlight}  2.7 & heterogeneous & 70\% & up to 49 & 0.67 & 6.09 & 3.09 \\ 
\rowcolor{highlight}  2.7 & heterogeneous & 70\% & up to 59 & 0.67 & 4.54 & 4.56 \\ 
\rowcolor{highlight}  2.7 & heterogeneous & 70\% & up to 69 & 0.67 & 5.68 & 5.82 \\ 
\rowcolor{highlight}  2.7 & heterogeneous & 70\% & up to 39 & 1 & 3.89 & 4.78 \\ 
\rowcolor{highlight}  2.7 & heterogeneous & 70\% & up to 49 & 1 & 4.96 & 7.01 \\ 
\rowcolor{highlight}  2.7 & heterogeneous & 70\% & up to 59 & 1 & 6.66 & 10.42 \\ 
  2.7 & heterogeneous & 70\% & up to 69 & 1 & 9.20 & 14.55 \\ 
  \hline
\rowcolor{highlight}   2.7 & heterogeneous & 50\% & up to 39 & 0.67 & 6.11 & 5.81 \\ 
\rowcolor{highlight}   2.7 & heterogeneous & 50\% & up to 49 & 0.67 & 6.70 & 7.13 \\ 
  2.7 & heterogeneous & 50\% & up to 59 & 0.67 & 7.48 & 8.52 \\ 
  2.7 & heterogeneous & 50\% & up to 69 & 0.67 & 8.35 & 9.62 \\ 
\rowcolor{highlight}   2.7 & heterogeneous & 50\% & up to 39 & 1 & 7.09 & 9.00 \\ 
  2.7 & heterogeneous & 50\% & up to 49 & 1 & 7.92 & 11.28 \\ 
  2.7 & heterogeneous & 50\% & up to 59 & 1 & 9.12 & 14.05 \\ 
  2.7 & heterogeneous & 50\% & up to 69 & 1 & 10.92 & 17.07 \\ 
  \hline
3 & homogeneous & optimal & NA & NA & 8.02 & 6.83 \\ 
\hline
  3 & heterogeneous & 90\% & up to 39 & 0.67 & 12.80 & 13.21 \\ 
  3 & heterogeneous & 90\% & up to 49 & 0.67 & 10.52 & 7.87 \\ 
  3 & heterogeneous & 90\% & up to 59 & 0.67 & 8.43 & 3.70 \\ 
  3 & heterogeneous & 90\% & up to 69 & 0.67 & 8.85 & 3.51 \\ 
\rowcolor{highlight}   3 & heterogeneous & 90\% & up to 39 & 1 & 6.13 & 1.74 \\ 
\rowcolor{highlight}   3 & heterogeneous & 90\% & up to 49 & 1 & 2.21 & 3.22 \\ 
\rowcolor{highlight}   3 & heterogeneous & 90\% & up to 59 & 1 & 4.23 & 6.95 \\ 
\rowcolor{highlight}   3 & heterogeneous & 90\% & up to 69 & 1 & 7.46 & 12.05 \\ 
  \hline
  3 & heterogeneous & 70\% & up to 39 & 0.67 & 10.78 & 6.29 \\ 
  3 & heterogeneous & 70\% & up to 49 & 0.67 & 8.03 & 3.09 \\ 
\rowcolor{highlight}   3 & heterogeneous & 70\% & up to 59 & 0.67 & 6.00 & 4.56 \\ 
\rowcolor{highlight}   3 & heterogeneous & 70\% & up to 69 & 0.67 & 6.57 & 5.82 \\ 
\rowcolor{highlight}   3 & heterogeneous & 70\% & up to 39 & 1 & 3.89 & 4.78 \\ 
\rowcolor{highlight}   3 & heterogeneous & 70\% & up to 49 & 1 & 4.96 & 7.01 \\ 
\rowcolor{highlight}   3 & heterogeneous & 70\% & up to 59 & 1 & 6.66 & 10.42 \\ 
  3 & heterogeneous & 70\% & up to 69 & 1 & 9.20 & 14.55 \\ 
  \hline
\rowcolor{highlight}   3 & heterogeneous & 50\% & up to 39 & 0.67 & 6.76 & 5.81 \\ 
\rowcolor{highlight}   3 & heterogeneous & 50\% & up to 49 & 0.67 & 6.70 & 7.13 \\ 
\rowcolor{highlight}   3 & heterogeneous & 50\% & up to 59 & 0.67 & 7.48 & 8.52 \\ 
  3 & heterogeneous & 50\% & up to 69 & 0.67 & 8.35 & 9.62 \\ 
\rowcolor{highlight}  3 & heterogeneous & 50\% & up to 39 & 1 & 7.09 & 9.00 \\ 
\rowcolor{highlight}   3 & heterogeneous & 50\% & up to 49 & 1 & 7.92 & 11.28 \\ 
  3 & heterogeneous & 50\% & up to 59 & 1 & 9.12 & 14.05 \\ 
  3 & heterogeneous & 50\% & up to 69 & 1 & 10.92 & 17.07 \\ 
 
\hline
\end{tabular}
    \caption{Mortalities and ICU admissions by scenario.  Like homogeneous strategies, heterogeneous strategies can perform suboptimally because they are too relaxed and allow too large of an epidemic in the mitigation period, or too strict and invite a second wave in the hypothetical resumption normal transmission levels.  Highlighted rows are scenarios which have lower mortalities than the optimal homogeneous measures for the same $R_0$.}
    \label{tab:scenarios}
\end{table}

\end{document}